# An Adaptive Load Mangement System using Predictive Control Optimization

Muneer Mohammad, PhD, IEEE member

*Abstract*—In this paper an adaptive load management system that uses predictive control optimization is introduced. This price elastic system is able to optimize the consumption of power and is fully autonomous and responsive to market clearing prices. The area of application chosen was an air-conditioning system that allows the end user to select a comfort zone that serves as the boundary conditions for the optimization algorithm. The temperature function that governs our algorithm is also derived and tested. Numerical examples are then presented to show the effectiveness of this system on day-ahead and real-time data from ISOs.

The developed system showed promising results and savings that will improve the utilization of present energy resources. Finally, the implementation of this system was discussed and some preliminary modeling was performed to show the potential realization of such a system.

## I. INTRODUCTION

Responsive load is a new idea that was presented with the introduction of the future smart grid. This concept tackles the complications introduced by the volatility of the demand function and the challenges of integrating intermittent energy sources with the current bulk power system. In [1], the smart grid solution of resolving these issues is presented. In [2], a model is proposed to determine the power consumption schedules that compromise between the total powers consumed and the preferences of the individual consumers based on the total cost and the level of quality of the service.

A model predictive economic dispatch method which integrates intermittent resources with price responsive loads is investigated in [3]. The system that was presented showed how load management could reduce total generation cost by allowing model predictive control to manage power consumption.

In [4], a multi-layered adaptive load management (ALM) system capable of integrating large scale demand response efficiently and reliably was proposed. In this paper, an autonomous, continuous optimization system is developed that would complement the ALM system. Numerical examples are then shown to illustrate the effectiveness of such a system. This system is relatively simple to implement on large scale and would greatly increase the efficiency of the power market.

The system developed is designed for the specific example of an air-conditioning (AC) system. The power consumed by the AC is efficiently managed through a circuit board that delivers the output of the load management system. Fig. 1. Shows the flow chart that maps the logic behind the implementation of the overall system. An efficient algorithm is then integrated into the optimization system to ensure resourceful results are obtained. Moreover, this process is designed to update every hour in a continuous fashion. This update feature ensures that the output relayed to the AC system is coherent with the cleared market price. Fig.1. shows the comfort zone feature that the user is able to control and which will control the boundaries of the optimization. Moreover, a "login code" (attached) will allow the user to activate the system operation.

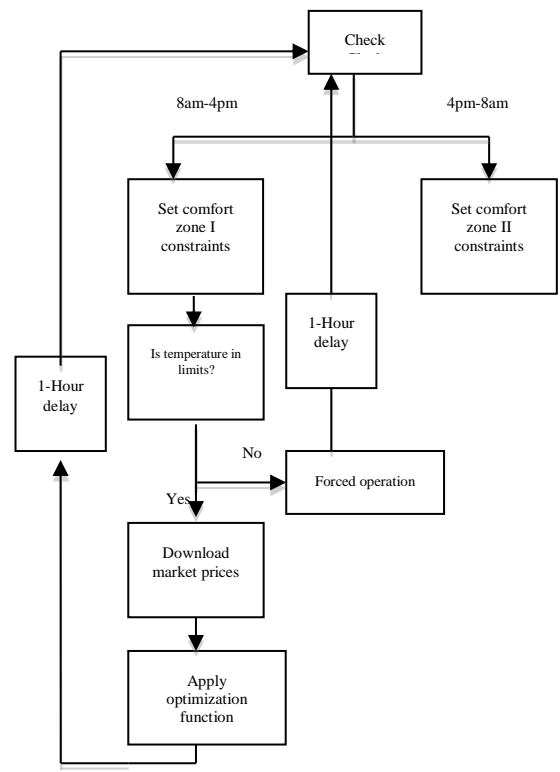

Fig. 1: Default System Operation

## II. ADAPTIVE LOAD MANAGEMENT WITH OPTIMIZATION CAPABILITY

In this section the adaptive load management system along with its optimization algorithm is introduced. The theory behind the derived temperature function and the optimization problem are also discussed. Fig. 2. shows the flow chart of the application of the algorithm and the steps that are involved in the information flow. The initial information about predicted prices are downloaded directly from the local Independent Operating System (ISO). This information is then utilized by the system until it outputs the optimized power output to the AC controls.

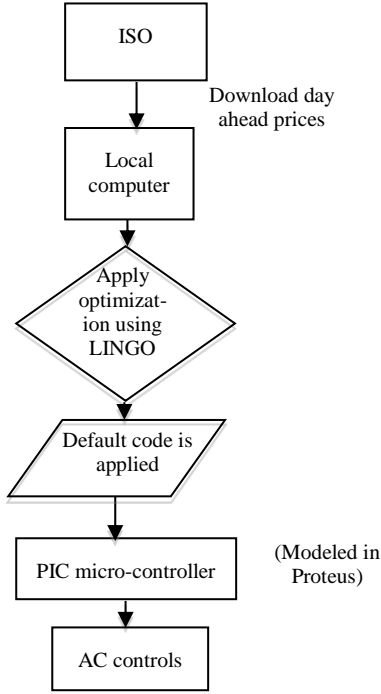

Fig. 2: Flow Chart of the Application of the Optimization Algorithm

The comfort zone settings set the constraints on the optimization problem; this comfort zone is set by the end-user of the AC system. Also, the capacity of the load (AC) and the ramping rate of the system further constrains the problem. However, the temperature function is what governs the temperature variation in the system. The temperature function is what couples the power consumed by the AC with the temperature of the zone.

The purpose of the login code is to allow the user to control this optimization of the load (AC). By using this program, the user is able to allow the system to operate under default conditions or to introduce the optimized control.

### A. Problem Formulation

The problem of interest in this paper is to determine the optimum load power consumption such that comfort zone constraints are not violated while minimizing total cost. We first review the problem formulation when the load is not responsive to electricity prices.

The predictive control model used in this paper is a finite optimal control problem solved on a future interval using predicted future variables. The initial state is used as the measured data of the current state of the actual system. The entire control model is optimized on an interval that is executed one step at a time.

The following notation is used throughout the paper:

C : cost of operating load over the optimization period;
$P_G$ : power consumption of load at time step (kW);
$P_G^{min}$ : minimum power consumption of load;
$P_G^{max}$ : maximum power consumption of load;
$\hat{h}$ : predicted market price (cents/kWh);
T : temperature at time step k;
$T^{max}$ : maximum temperature limit;
$T^{min}$ : minimum temperature limit;
$T_{out}$ : outdoor temperature (°F)
$T_{set}$ : temperature set by user;
k : number of samples in the optimization period;
ε : factor of inertia;
γ : thermal conversion efficiency;

### B. Inelastic Load Power Consumption

In this paper, we assume that the expected price is given for the next 24 hours. Therefore, the end user presets the comfort limits over a certain day-ahead interval. The thermodynamics of the end-user's temperature control is modeled in relation to the energy usage over each individual step [6].

For inelastic loads, the temperature restrictions are determined by the user-specific set point. This inelasticity in the power requirement increases the cost of operation. The set point governs the load behavior, and the limits on temperature become very strict. This model represents the load behavior of current systems. Also, the cost is dependent on the power consumed, and there is no integration of the predicted price. Therefore, the predictive optimization model of the inelastic demand can be formulated as follows:

Solve: $\min_{P_G} \sum_{k=1}^{24} (C(P_G(k)))$

s.t. $T(k+1) = T(k) = T_{set}$
$0 \leq P_G(k) \leq 20$
$70 \leq T(k) \leq 75$

### C. Derivation of Temperature Function

The derivation of the temperature function is a critical step because, this function is what couples the temperature of the zone with the power consumed by the AC system. Also, the mathematical model that describes the temperature at the next time step needs to be adaptive and responsive to external factors. Thus, the outside temperature and the factor of inertia need to be introduced to our formulation.

Therefore, we let $T(t)$ to be the temperature inside the controlled zone at time $t$. Then the rate of change of the temperature inside the controlled zone could be modeled by the following:

$$\frac{dT(t)}{dt} = K_1(T_{Out} - T(t)) - K_2(T(t) - T_d)$$

## D. Predictive Model Optimization for Price-Responsive Loads

With price responsive load, the optimization model problem now has predicted prices as a variable. As a result, the objective function becomes dependent on the power consumed by the load as well as the predicted price. Moreover, the temperature model is more dynamic and depends on the power consumed. Temperature constraints are now within a comfort zone determined by the user. The overall problem formulation is as follows:

$$\text{Solve: } \min_{P_G} \sum_{k=1}^{24} (C(P_G(k), \hat{h})$$
$$\text{s.t. } T(k+1) = \varepsilon T(k) + (1-\varepsilon)(T_{out} + \gamma P_G)$$
$$0 \leq P_G(k) \leq 20$$
$$70 \leq T(k) \leq 75$$

In the above formulation, the load is considered to be responsive to the market clearing price. The elastic loads can follow the temporal variations of the cleared price cleared. Energy usage can be optimized with such predictive control of the load. The optimization is done in a window method, where the optimization is done over a 24-step interval. Based on this formulation, a numerical example is presented the next section to illustrate the effectiveness of the formulation.

## III. NUMERICAL EXAMPLES ILLUSTRATING THE SAVINGS OF PREDICTIVE CONTROL OPTIMIZATION

In this section, numerical examples of the predictive control optimization are shown on an air-conditioning load. The load specifications are given in Table I. Two scenarios are simulated for comparison purposes. The optimization predictive control is executed for a 24-hour future window. In the first scenario, an inelastic load is assumed. In the second scenario, the load is assumed to be price-responsive. Also, the outside temperature is assumed to be a constant function across the time window. The simulation is implemented in Microsoft Excel 2007 with a link to LINGO 12.0 for the optimization operation. Visual Basic is also utilized to automate the predicted price-acquiring process from ERCOT website.

Fig. 3. shows the comparison of power consumption by the air-conditioning unit in the inelastic model and in the elastic (price-responsive) model. In the elastic load scenario, the total cost was calculated to be $4197.2, while the cost of the inelastic load was $6915.9. This difference is due to the flexibility of the power consumed by the load in the elastic scenario. Since the predictive model tends to increase consumption in time intervals where prices are low and decrease consumption in intervals where prices are high, while satisfying the comfort zone limits. Fig. 4. shows the temperature variations of both scenarios. In the inelastic scenario, the temperature is governed by the user-set point and is maintained at that level. On the other hand, in the elastic load model the user identifies a comfort zone that dictates the limits of the temperature. The temperature generated by the elastic model satisfies the comfort zone constraints and simultaneously decreases cost of power consumption. Fig. 5, shows the behavior of the power consumed when compared with the predicted price. This comparison is important because it clearly shows that as the market-cleared price increases, the predictive control works to compensate this effect on the cost. This is done by increasing load powers prior to the price increase, resulting in the ability to decrease load powers during price peaks. This mechanism of predictive control has the ability to maintain comfort zone limits while reducing cost. As illustrated, by introducing an optimizing predictive model of the control for our load consumption, the cost of operation could be reduced.

TABLE I
LOAD SPECIFICATIONS OF THE AIR-CONDITIONING SYSTEM

| | |
|---|---|
| Cooling System Capacity | 80,000 Btu/Hr (ARI Rating) / 20.0 KW |
| Evaporator Air Flow Capacity | 2354 CFM / 4000 m3/Hr |
| Unloading Steps | 100% - 50% |
| Factor of Inertia | 0.8 |
| Drive Ration | 1:1 |
| Total Power Input | @12VDC 110A / @24VDC 70A |
| Controls | Microprocessor |

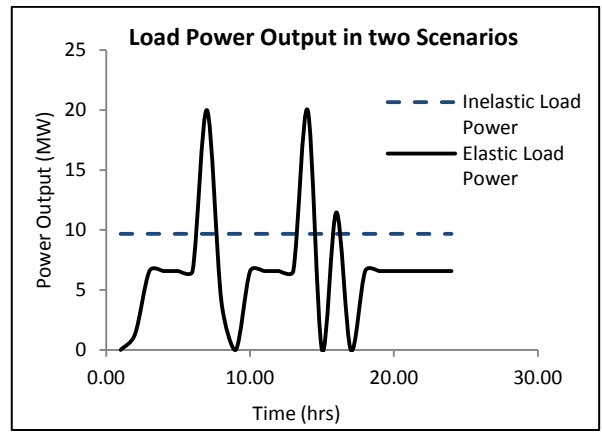

Fig. 3. Load Power Output in the two scenarios

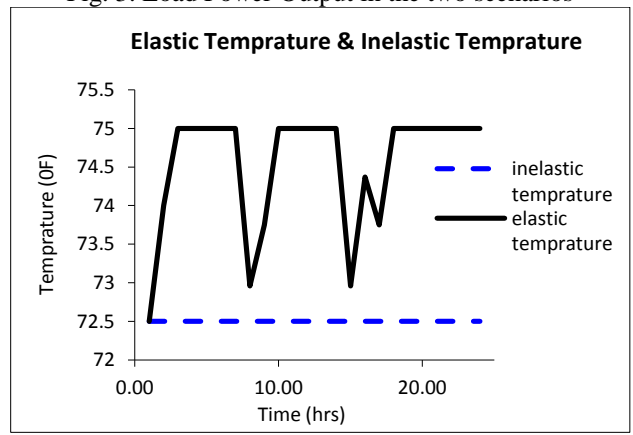

Fig.4. Elastic and Inelastic Temperature in the two scenarios

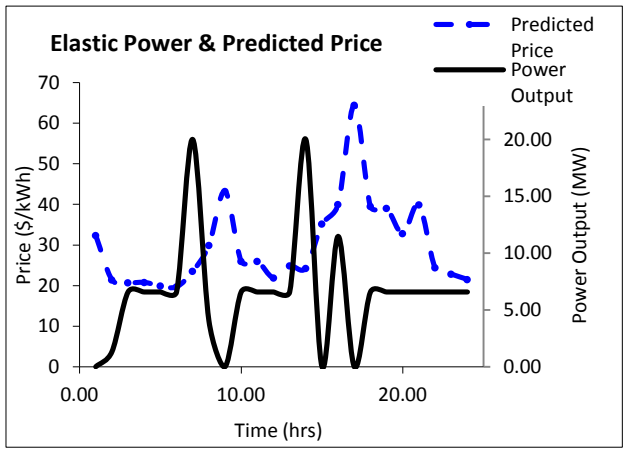

Fig. 5. Elastic Power & Predicted Prices for the look-ahead window

This predictive control system was also applied to a different scenario to study the effects of inaccurate prediction. Locational marginal prices were obtained from the ISO New England website, to complete this study. The look-ahead optimization was carried out with the day-ahead LMP's and the solutions were examined. These results were then compared to the output when the real time LMP's were used. Figures 6 & 7 show the results obtained. The temperature variation was not signification and greatly resembled the expected temperature.

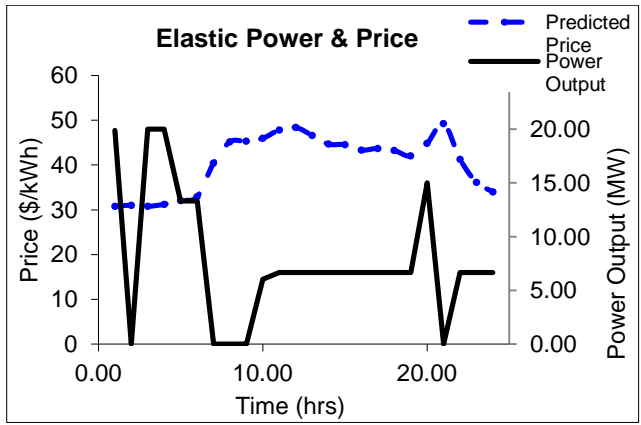

Fig. 6. Optimization Power and Day-ahead Prices

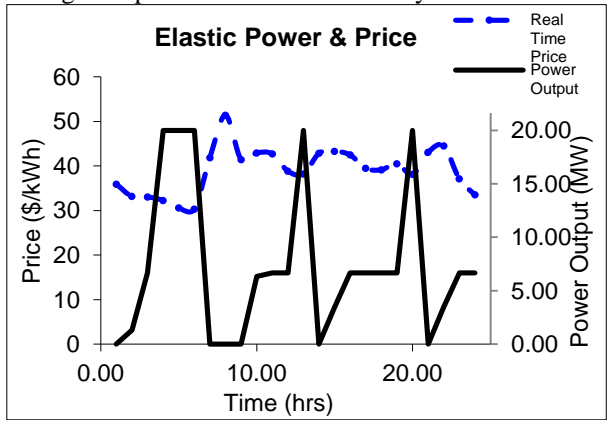

Fig. 7. Optimization Power and Real-time Prices

From the figures we can see that the real time LMPs are much more volatile and changing. This affects the optimization output and the potential savings. Further study investigates the amount of savings projected versus the amount of savings according to the real time market prices. The projected cost according to the predictive optimization is $7109; however, when the real time market prices were used the cost was $6879. This shows that for this particular case the optimization system works very efficiently and is able to provide savings higher than expected according to the real time LMPs.

Further simulations were done to test the efficiency of this system when applied to a real situation where on-line optimization is required. In the power market day-ahead prices, all the prices are predicted and the only known price is the LMP for the current hour. Therefore, at the beginning of the window of optimization the only price certain is the current price or in other words the price at hour zero. As a result, the predicted power output of the load has to be updated continuously with the market clearing prices on the hour. This system was applied to a whole window of iterative simulations over 24 hours, where the optimization was run with new market prices at the new hour. The results showed that the optimization function almost maintains a non-changing behavior; however, in some instants the predicted prices differed greatly than the actual price, which caused an slight change in the behavior of the predicted power output. Fig.8. shows two consecutive power outputs at hours & that show how the power output remains constant. While Fig. 9. shows an example of how error in predicting prices could cause a fluctuation in the predicted power of the A.C.

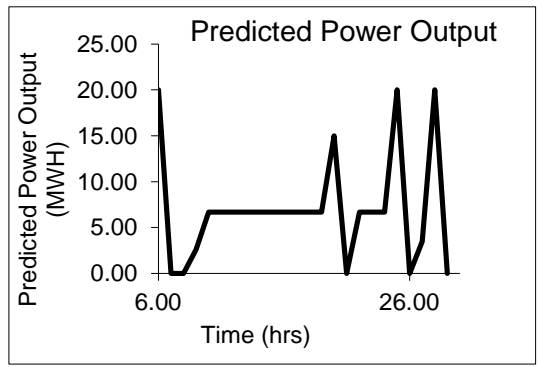

Fig.8a. Predicted Power output at k=6

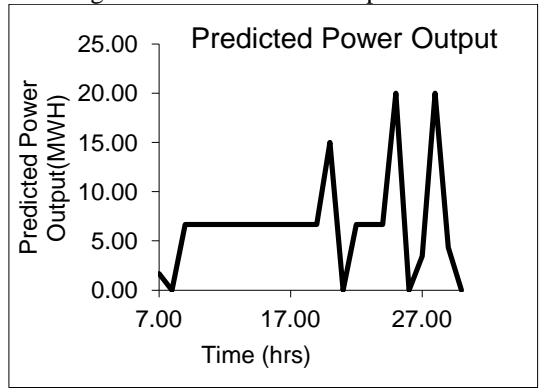

Fig. 8b. Predicted Power Output at k=7

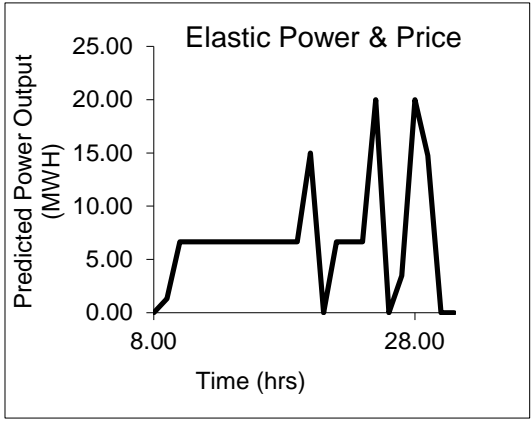

Fig. 9a. Predicted Power Output at k=8

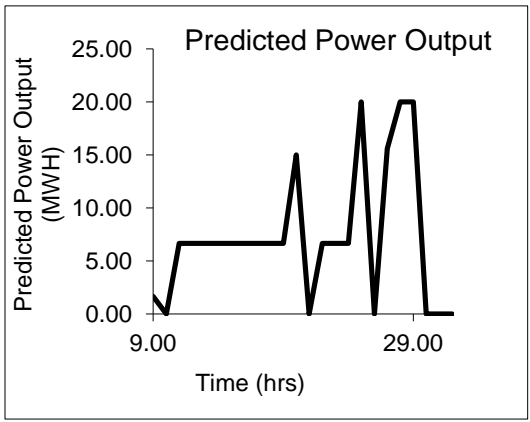

Fig. 9b. Predicted Power Output at k=9

From Figures 8 & 9 above the effect of error of price prediction could be studied. In Fig. 8. the output of the optimization function remained constant even after updating the real market price at hour 7. On the other hand in Fig. 9. the output function had some abrupt changes specially at hour 29. The optimization system in the previous time step required that at hour 29 the A.C. unit be operating at around 15MWH. However, when price for hour 9 was updated the optimization system required the A.C. to operate at full capacity at hour 29. This comparison shows the slight difference that is caused by the error in predicting market prices. Nevertheless, this predictive optimization system proved to be an efficient and robust system that is capable of reducing cost of operation of a load without sacrificing end-user comfort.

## IV. HARDWARE IMPLEMENTATION AND INTERFACE CIRCUIT

In this section, the hardware implementation of our system design is presented and discussed. The realization of such a system was designed on Proteus. The main component that was used in our design was a PIC microcontroller which can be obtained in different packages, such as a conventional 40-pin DIP (Dual In-Line Package), square surface mount or socket format [5, 6]. The PIC microcontroller architecture has been introduced in [6].The source code is written on a PC host using a C compiler, assembled and downloaded to the chip as shown in Fig. 8.

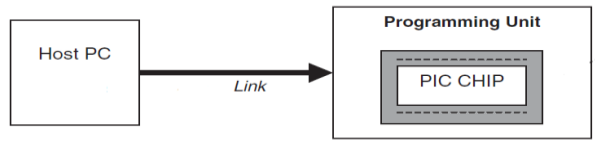

Fig. 8. Programming Operations of the PIC

A serial link is the connection unit that serves as a delivery communication between the host PC and the programming unit. The PC runs a C-compiler, so that once a program has been written and assembled, it is downloaded by placing the chip in a PICSTART Plus connected to a PC. This unit is available in the instrumentation room in ZACH 100M.

After the optimization is executed in the Lingo and Excel software, the objective of the PIC is to extract the proposed power consumption information from the Excel sheets and then deliver the information to the AC controls. The output of the PIC controller is a twenty level representation of the power outputted by the optimization control. The schematic circuit shown in Fig. 10. describes the components used to design a circuit that will be used in order to connect it to the AC when it converts into a real application. We will connect the output to 20 LEDs to represent the actual illumination within a specific period.. Note that the output changes as the input changes. This work can be modified to include any delays by modifying the "FOR loop" used to make it hourly.

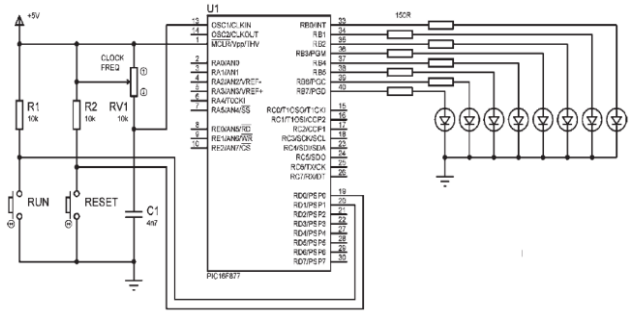

Fig. 10. Components of the Circuit Design

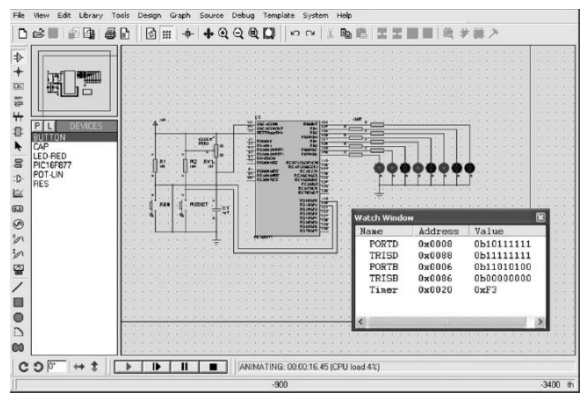

Fig. 11 Proteus Simulation of the Circuit Design

In our design, a 20 MHz crystal circuit is used to give a 200ns instruction execution time and to create a precise frequency when subject to electrical stimulation, which is more convenient for analyzing program timing. Moreover, the crystal oscillator needs to be physically near the MCU to prevent the crystal from oscillating, or affecting the resonant frequency. In addition, a 10pF capacitor is used to avoid the clock signal causing a malfunction, and two 15Pf are used to stabilize the frequency.

As shown in Fig. 11, the Proteus simulation software was used to write in C-compiler in order to provide virtual instruments which can be used to build up the circuit, just as in a real application. In short, we can summarize the steps of the designed circuit operation in the following manner: At the beginning, the user enters the Uniform Resource Locator (URL) of the requested site on the Excel sheet, and then the optimization is processed by using Lingo software, the default code then converts the entered text to a binary request and sends it to the PIC controller.

The most challenging task in this project is designing an interface circuit that meets all the specifications mentioned in previous sections. However, building such circuit requires an intensive study. A simple method of controlling AC Conditioning is to use a relay as a switch to control and to send the signal with certain delays based on the optimization; however, in practice there are some tricky issues associated with controlling inductive loads